\documentclass[12pt,reqno]{amsart}

\usepackage{amsxtra}
\usepackage{amsfonts}
\usepackage{amsmath,amssymb,stmaryrd} 
\usepackage[english]{babel} 
\usepackage{graphicx} 
\usepackage{mathrsfs} 
\usepackage{layout}
\usepackage{color}

\newcommand{\al}{\alpha}

\newcommand{\ome}{\omega}

\newcommand{\ba}{\begin{eqnarray}} 
\newcommand{\ea}{\end{eqnarray}}
\newcommand{\be}{\begin{equation}}
 \newcommand{\ee}{\end{equation}}
\newcommand{\bdm}{\begin{displaymath}}
\newcommand{\edm}{\end{displaymath}} 
\newcommand{\brr}{\begin{array}}
\newcommand{\err}{\end{array}}

\newcommand{\bml}{\begin{gather}} \newcommand{\eml}{\end{gather}}

\newtheorem{teo}{Theorem}[section]
\newtheorem{prop}{Proposition}[section]
\newtheorem{rem}{Remark}[section]
\newtheorem{lemma}{Lemma}[section]

\newcommand{\spaz}{\vspace{.5cm} \noindent}

\newenvironment{dem}{\vspace{.2cm}\noindent {\bf Proof }\\}{\newline \spaz \hspace{1cm} \hfill $\square$ \newline}

\numberwithin{equation}{section}

\newcommand{\R}{{\mathbb R}}         
\newcommand{\var}{\varepsilon}
\newcommand{\ud}{\mathrm{d}}
\newcommand{\pa}{\partial}

\begin{document}
\begin{center}
{\large\textbf{Strong Semiclassical Approximation of Wigner Functions for the Hartree Dynamics }}
\end{center}
\vspace{0.5cm}
\begin{center}
A. Athanassoulis\footnote{CMLS UMR 7640, \'Ecole polytechnique, Palaiseau, France  \\ \emph{e-mail:} agis.athanassoulis@math.polytechnique.fr}, 
T. Paul\footnote{CNRS and CMLS UMR 7640, \'Ecole polytechnique, Palaiseau, France  \\ \emph{e-mail:} thierry.paul@math.polytechnique.fr},
F. Pezzotti\footnote{Departamento de Matem\'aticas, Universidad del Pa\'is Vasco, Spain  \\ \emph{e-mail:} federica.pezzotti@ehu.es},
M. Pulvirenti\footnote{Dipartimento di Matematica ``G. Castelnuovo'', Universit$\grave{\text{a}}$ di Roma ``La Sapienza'', Italy \\  \emph{e-mail:} pulvirenti@mat.uniroma1.it }
\end{center}
\vspace{0.5cm}
\begin{center}
\textbf{Abstract}
\end{center}

{\footnotesize We consider the Wigner equation corresponding
to a nonlinear Schr\"odinger evolution of the Hartree type in the  
semiclassical limit $\hbar\to 0$. 

Under appropriate assumptions on the initial data and the interaction potential, we show that the Wigner function is close in $L^2$ to its weak limit, the solution of the corresponding Vlasov equation. The strong approximation allows the construction of semiclassical operator-valued observables, approximating their quantum counterparts in Hilbert-Schmidt topology.

The proof makes use of a pointwise-positivity manipulation, which seems necessary in working with the $L^2$ norm and the precise form of the nonlinearity. We employ the Husimi function as a pivot between the classical probability density and the Wigner function, which -- as it is well known -- is not pointwise positive in general.

}

\tableofcontents

\section{Introduction}\label{introo}

\subsection{Formulations of the problem}

The time evolution of a density matrix $A^\var(t)$ in a self-consistent field of  Hartree type  is described by the  initial value problem
\begin{equation}
\label{Heisemberg}
\left\{
   \begin{aligned}
   &i\var \pa_t A^\var (t)=[H^\var,A^\var (t)],\\
   & A^\var(0)=A_0^\var,   \end{aligned}
   \right.
 \end{equation}  
where $[\cdot. \cdot ]$ denotes the commutator, $\var:=\hbar$  is the Planck constant and  $A_0^\var$ is the initial datum. The dimension of the system is chosen equal to three. 

We will be interested in initial data of the mixed state form; see the statement of Theorem \ref{MAINteoState} for details. We are going to look at the dynamics (\ref{Heisemberg}) in the regime $\var \ll 1$, the so-called semiclassical regime, where we expect the dynamics to ``approach'' the one of the corresponding classical system. The Hamiltonian $H^\var$ for a particle of mass $m$ is given by
\begin{equation}
H^\var=-\frac {\var^2}{2m}Ê\Delta +V
 \end{equation} 
and $V$ is a self-consistent potential determined by
\begin{equation}
V(x,t)=\int_{\R^3}\phi(x-x')\rho_t^\var (x')\ud x',
\end{equation}
where $\phi(x)$ is the two-body interaction and $\rho_t^\var (x)=\rho^\var (x,x,t)$ is the position density given in terms of  the integral kernel $\rho^\var(x,y,t)$ of the density matrix $A^\var(t)$ (see below). The pair interaction potential $\phi$ is assumed to be spherically symmetric, a natural assumption from the physical point of view. 

It will be shown that the Wigner function corresponding to the operator $A^\var$ (see below) converges in $L^2$ to the probability distribution $g=g(x,k,t)$ that solves the corresponding classical Vlasov equation:
\begin{eqnarray}\label{eq: 21}
\pa_t g\ + k\cdot \pa_x g = T_0^g g,
\end{eqnarray}
where
\begin{eqnarray}\label{eq: 22}
&&\left( T_0^g g\right)(x,k,t)= \left(\partial_x \phi\star \rho_t^g\right)\cdot \partial_k g(x,k,t) = \left(\int_{\R^3} \ud y\ \partial_x \phi(x-y)\rho_t^g(y)\right)\cdot \partial_k g(x,k,t),\nonumber\\
&&
\end{eqnarray}
and $\rho_t^g=\rho_t^g(x)$ is the spatial probability density associated with $g(x,k,t)$, namely:
\begin{eqnarray}\label{eq: 23}
\rho_t^g(x)= \int_{\R^3} \ud k\ g(x,k,t).
\end{eqnarray}
Moreover, a semi-classical approximation of $A^\var$ will be constructed; the precise statement of the result is given in Section \ref{ssst}.

\vskip 0.4cm
Problem (\ref{Heisemberg}) has several equivalent formulations, each of them yielding a corresponding initial value problem. 
For example, by looking at the integral kernel $\rho^\var(x,y,t)$ defined by
\begin{equation}
(A^\var(t)\, G)(x) = \int{\rho^\var(x,y,t)G(y)dy},\ \ \ \ \ \ \ \ \ \ \ \ \ \ \ \ \ \ \ \ \ 	\ (\text{for any} \ G\in L^2(\R^3))
\end{equation}
it is easy to check that it satisfies equation (\ref{VonNeu}) -- sometimes called von Neumann equation.
Another description is in terms of the Wigner function $f^\var(x,k,t)$ which is related to 
the density matrix $A^\var(t)$ through the following (\emph{Weyl}) transformation:
\begin{equation}
(A^\var(t)\,  G)(x)=\int{ f^\var(\frac{x+y}{2},k,t)e^{i(x-y)\frac{k}{\var}}G(y)dkdy}, \ \ \ \ \ \ \ \ \ (\text{for any} \ G\in L^2(\R^3))
\end{equation}
(see equations (\ref{Weyl}),  (\ref{eqeagmal}) for the relation between the Wigner function and the Weyl symbol). 
The Wigner function satisfies equation (\ref{WignerEq}) -- sometimes called Wigner equation.
A very straightforward connection between the three descriptions is that
\begin{equation}
\label{trhssomat}
||A^\var||_{HS}=||\rho^\var(\cdot, \cdot)||_{L^2(\R^3\times\R^3)}=
(2\pi\var)^{\frac{3}2}
||f^\var||_{L^2(\R^3\times\R^3)},
\end{equation}
where $\left\| \cdot \right\|_{HS}$ is the Hilbert-Schmidt norm.
This in particular allows to translate easily $L^2$ estimates between the different formulations, and to transfer approximations from the function level to the operator level.

It must be noted that a natural assumption for the initial datum in the Density Matrix Formalism is that the operator is positive semi-definite and trace-class (with trace equal to one), i.e. it has a singular value decomposition of the form
\begin{equation}
\label{eqSVD}
A^\var_0=\sum \lambda_m |u_m\rangle \langle u_m|
\end{equation}
with $||u_m||_{L^2(\R^3)}=1$, $\lambda_m \geqslant 0$, $\sum \lambda_m =1$. It is well known that the trace is preserved in time. This is an important physical fact, as the trace is the quantum counterpart of the total probability of a classical density in a statistical formulation.
The trace is given by $tr(A^\var)=\int{\rho^\var(x,x)dx}=\int{f^\var(x,k)dxdk}$ (see also Lemma \ref{lml1reg} and the discussion below).

\subsection{Physical context}
Eq.n (\ref{Heisemberg}) describes the situation in which we have a large number of particles in a mean-field regime (see for instance \cite{SPOHN,HEPP,GV}, for the case of smooth potentials, and \cite{BGM,EY,BEGMY,ErdBenji,KnowlesPICKL,Pickl} for more singular interactions).
In this paper we want to study the semiclassical behavior (for $\var \to 0$) of the solution of equation (\ref{Heisemberg}).

\vskip 0.35cm

The Wigner function is a well known tool in the study of the semiclassical limit of quantum dynamics
(see definitions (\ref{pure}) and (\ref{mixed}) below). 
Indeed, there are many works using the Wigner function to study the semiclassical limit of a number of problems (linear, non-linear,
 stochastic, 
systems etc) see e.g.  \cite{MarkowichN,Markowich,Paul,GMMP,PPpaper,PULVIRENTI,RPK} for a very small selection, and the references therein.
One of the main advantages, is that the (formal, at this level) limit, as $\var \to 0$, of the Wigner  equation is typically some familiar equation of classical statistical mechanics. In that sense, the ``correspondence principle'' between classical and quantum mechanics is quantified in a straightforward, easy to present way. Indeed, for the problem we deal with here, for example, guessing the result from formal calculations is pretty straightforward.

In most of the existing literature, the notion of convergence is in weak topology (see e.g. the works mentioned above for precise statements). Indeed, the weak-$*$ semiclassical limit for this problem is worked out e.g. in \cite{Paul}. In fact (outside coherent states techniques), until very recently, virtually all the results were in weak topology. There is a natural analytical question of understanding when and why convergence in some natural strong topology fails; 
moreover, if one has possible numerical applications in mind, it would be desirable to know e.g. whether 
(possibly large in $L^2$ or pointwise sense) oscillations develop or not. Quantifying constructively the rate of convergence in terms of the data of the problem is also another natural question.

Another big family of methods that yield strong topology semiclassical asymptotics (in linear as well as nonlinear problems) is based on coherent states (e.g. \cite{Carles,HAG1,HAG,HEPP,PaulU,russians,ours}). However, this is not really pertinent here, as coherent and mixed states (the kind of data we treat here) in nonlinear problems behave quite differently.

\vskip 0.2cm
In the quantum-classical correspondence, the idea is that the Wigner function converges, in the semiclassical limit, to a classical phase-space probability measure. However, the Wigner function itself is not pointwise positive in general.
Working around this fact will be among the main points of the proofs.
Indeed, it has often been remarked that the extensive arsenal of positivity techniques, developed in the context of classical phase-space equations, would be a good ingredient to transfer to their quantum counterparts. This becomes even more important in non-linear problems. In this paper we employ such pointwise-positivity techniques to the (nonlinear) Wigner equation, for the first time to the best of our knowledge. The key idea is to work with the Husimi function, a variation of the Wigner function, which does translate the operator positivity into pointwise positivity on the phase-space. The equation for the Husimi function itself has been derived only recently in closed form, and it is of infinite order in general \cite{AMP}; 
it helps us in guessing the precise manipulations that are needed here. However, once formulated, the estimates we need can be proven without using the infinite-order Husimi equation itself. It should also be noted that the Husimi function is used only in the proof; it does not appear in the statement of the result, which is formulated in terms of the Wigner function. We believe that such use of positivity techniques in nonlinear Wigner equations could provide a fruitful approach in other problems as well.
\subsection{Notation}

We specify that here and henceforth we use the following conventions for the Fourier transform in $\R^d$: 
\begin{equation}\label{Fourier}
\widehat{G}(k)=\mathcal{F}_{x \rightarrow k}[G(x)]=
\int_{\R^d}e^{- i k x}\ G(x)\, \ud x.
\end{equation}

\vskip 0.6cm
The Weyl Quantization is defined as follows: for any  $F\in L^2(\R^3\times\R^3)$  and $G\in L^2(\R^3)$,
\begin{equation}\label{Weyl}
(Op^{Weyl}(F)G)(x)=\var^{-3}\int{ F\left(\frac{x+y}{2},k\right) e^{i(x-y)\frac{k}{\var}}G(y)dkdy }.
\end{equation}

\vskip 0.6cm
We denote by $H^\nu(\R^d)$ the Sobolev space $W^{\nu,2}(\R^d)$ of functions in $L^2(\R^d)$ whose derivatives 
up to the order $\nu$ are also in $L^2(\R^d)$, i.e., for any function $G$ on $\R^d$
\begin{equation}
||G||_{H^\nu(\mathbb{R}^d)}=\sum\limits_{|a|=0}^\nu ||\partial^a_x G(x)||_{L^2(\mathbb{R}^d)}
\end{equation}

\vskip 0.6cm
We denote by $C_b^k(\mathbb{R}^d)$ the space of continuous and uniformly bounded functions whose derivatives, up to the order $k$, are also continuous and uniformly bounded.

\subsection{Organization of the paper}
The plan of the paper is the following. Section 2 is devoted to the statement of our main result
together with some remarks concerning it and fixing the notations. In Section 3 we recall the main features 
of the Hartree dynamics rephrased in the Wigner formalism. In Section 4 we introduce the definition and various 
properties of the Husimi Transform (HT).
Then, in Section 5 we recall the main features of the Vlasov evolution, namely, 
the classical dynamics we recover in the limit $\var\to 0$. 
 Finally, in Section 6 we prove the main result of this paper.

 \section{The main result}\label{mainn}
\subsection{Statement of the result}
\label{ssst}

All over the paper we make the following assumptions for the  interaction potential that we suppose to be spherically symmetric:
\begin{eqnarray}
&&\imath)\ \ \phi\in H^1(\R^3)\label{phiH1}\\
&&\imath\imath)\ \ \int_{\R^3} dS\, \vert\widehat{\phi}(S)\vert\, \vert S\vert^n<+\infty,\ \ \ n=0,1,\dots,4\label{MOM4}.
\end{eqnarray}

\begin{teo}\label{MAINteoState} Under the above assumptions on $\phi$, let $A^\var(t)$ be
the solution of the Hartree problem (\ref{Heisemberg}), and   $f^\var(t)$ the Wigner function associated with it. 
Denote $f^\var(0)=f^\var_0$, and let us suppose that:
\begin{itemize}
\item $\exists \ C, C'>0$ such that
\begin{equation}\label{hypf}
\begin{array}{c}
\left\| f_0^\var\right\|_{H^{3}(\R^3\times\R^3)} \leq C,\ \int \, f_0^\var(x,k)\, |k|^2\, dx d k \leq C'\  
 \mbox{ and }\int{f^\var_0(x,k)dxdk}=1.
 \end{array}
\end{equation}
\item
$\exists \ M_0>0$,  $\alpha \in (0,1]$ (independent of $\var$), such that
\begin{eqnarray}\label{truc}
\int_{|k|>\frac{M_0}{2}} |f^\var_0|^2 dxdk = O(\var^{2\alpha}),\label{truc21} \\
\int_{|k|>\frac{M_0}{4}} |f^\var_0| dxdk =  O(\var^\alpha),
\label{truc22}
\end{eqnarray}
Moreover
\begin{eqnarray}
||f^\var_0||_{L^1(\R^3)}= O(\var^{\alpha+\frac 3 2}e^{{M_0^2}/{16\var}}).\label{truc23}
\end{eqnarray}

\end{itemize}

\vskip 0.5cm

Let $\left\{g_0^\var\right\}_\var$ be any  family of probability distributions bounded in $H^{3}(\R^3\times\R^3)$ 
(uniformly in $\var$),
supported on the set $\{ (x,k)\in \R^3\times\R^3: \, \, \, \vert k\vert\leq  M_0\}$, such that

\begin{eqnarray}\label{fgclose}
\left\|f_0^\var-g^\var_0\right\|_{L^2(\R^3\times\R^3)}=  O(\var^{\alpha}),
\end{eqnarray}
(for example $g_0^\var=
\chi_{M_0}\widetilde f^{\varepsilon}_0/\int  \chi_{M_0}\widetilde f^{\varepsilon}_0dxdk$ 
where $\widetilde f^{\varepsilon}_0$ is the Husimi Transform of $f^{\varepsilon}_0$  as in (\ref{eqDefPhi}) 
and $\chi_{M_0}$ is a  smooth  function identically equal to $1$
for  $\vert k\vert\leq M_0/2$ and vanishing for $\vert k\vert\geq M_0$, see Proposition \ref{const} below).
\vskip 1.5cm
Then, if we denote by  $g^\var(t)$  the solution of the Vlasov equation (\ref{eq: 21}) with initial datum $g^\var(0)=g^\var_0$ 
, there exist positive constants
$C_0,C_1,C_2$ 
(else we need to say independent of $\var,\ t$), 
such that
\begin{equation}\label{MainWIGNER}
\left\| f^\var (t) -g^\var(t) \right\|_{L^2(\R^3\times\R^3)} \leqslant C_0\,  e^{C_1\, e^{C_2\, t}} \var^{\frac 2 7\alpha} .
 \end{equation}
 
\noindent In particular it follows that the  density matrix
$A^\var(t)$ can be approximated by the semiclassical operator $B^\var(t)$ whose Wigner function is $g^\var(t)$. 
More specifically
\begin{equation}\label{eq: teo01}
\frac{||A^\var(t)-B^\var(t)||_{HS}}{||A^\var(0)||_{HS}} \leqslant C_0\,  e^{C_1\, e^{C_2\, t}} \var^{\frac 2 7\alpha}  .
\end{equation}
\end{teo}

\vskip 0.5cm
 
In the statement of the theorem we do not require the existence of a semiclassical  limit $f^\var_0 
\to g_0^0$, we strictly need only assumptions on the quantum data $f^\var_0$. These assumptions
guarantee the existence of semi-classical initial data $g_0^\var$  fulfilling  the hypotheses of Theorem \ref{MAINteoState}, as established by
Proposition \ref{const} below.  
On the other hand, if one supposes a priori the existence of  such a family,
 the statement of Theorem \ref{MAINteoState} holds,  without assuming  conditions  (\ref{truc21}), (\ref{truc22}) and (\ref{truc23}) .
\begin{prop}\label{const}
Let $ \chi_{M_0}=\chi_{M_0}(|k|) $ be a monotone $C^{\infty}$ function satisfying 
$\chi_{M_0}=1$ if $|k|\leq\frac{M_0}{2}$ and $\chi_{M_0}=0$ if $|k|\geq M_0$, and let $\widetilde f^{\varepsilon}_0$ be the the Husimi Transform of $
f^{\varepsilon}_0 $ as in (\ref{eqDefPhi}).

Suppose  $g^\var_0$  be defined by
\begin{equation}\label{ansatz}
g^\var_0 =\frac{\chi_{M_0}\widetilde f^{\varepsilon}_0 \, }
{\int  \chi_{M_0} \widetilde f^{\varepsilon}_0dxdk},
\end{equation}
then
\begin{enumerate}
\item \label{enumitemappend1}
$\int{g^\var_0 dxdk}=1$ \mbox{and} $g^\var_0\geq 0$\\
\item \label{enumitemappend2}
$||g^\var_0||_{H^3}=O(1)$\\
\item \label{enumitemappend3}
$supp \, g^\var_0 \subseteq \{ |k|\leq M_0 \}$\\
\item \label{enumitemappend4}
$||f^\var_0-g^\var_0 ||_{L^2} = O(\var^{\alpha})$.
\end{enumerate}
\end{prop}
The proof of Proposition \ref{ansatz} is given in the Appendix.
\vskip 1.5cm

\noindent {\bf Remarks:}

\begin{itemize}

\item
For a sharper expression on the behaviour in time of the error see (the end of) the proof in Section \ref{setionproof765}. 
In this form, the constants  $C_0$, $C_1$ depend on
 the $H^3$-norm of the initial data $f_0^\var$ and $g^\var_0$, the initial total energy of $f^\var_0$, 
  suitable moments of  $\widehat{\phi}$ (as in equation (\ref{MOM4})), and $||\phi||_{H^1}$. All these quantities are bounded uniformly in $\var$ by assumption. 
The constant $C_2$ depends on those quantities mentioned above that involve only $\phi$.

\item
The investigation of the semiclassical limit of the Wigner Transform 
by looking at the $L^2$ asymptotics arises quite naturally because such a norm is invariant under the time evolution. However, while for the linear case everything goes on easily (provided that the potential is sufficiently smooth; see e.g. \cite{MarkowichN,PULVIRENTI,AP}), for the nonlinear case  one has to face an extra difficulty, which is the motivation of the present paper. In fact the $L^2$-norm of the difference between the Wigner Transform and its classical counterpart is estimated in terms of the $L^1$-norm of the same difference. Therefore, to conclude, we need a control of large momenta. This could be achieved by the energy conservation, but for an effective use of it we would need the positivity of the Wigner Transform, which is not the case. This difficulty has been overcome by using the Husimi Transform.
\item
For the sake of concreteness we work in dimension three, but our results hold as well in any dimensions.  We also expect the result to hold without any important 
differences if a regular enough external potential is added.
\end{itemize}

\subsection{Remarks on the Initial Data and Regularity Assumptions}
\label{sectassumptions00}
\begin{rem}
The assumptions we made on $\phi$ and $g^\var_0$ guarantee the existence and uniqueness for the Vlasov equation (\ref {eq: 21}) in the space of probability measures. 
\end{rem}
\begin{rem}\label{ex}

An explicit example for which the  assumptions of Theorem \ref{MAINteoState} are verified, is a superposition of coherent states, namely, for an 
$\var-$independent probability density $g_0 \in H^{3}(\R^3\times\R^3)$ and supported on $\{ (x,k)\in \R^3\times\R^3: \, \, \,  |k|\leq M_0 \}$,
\begin{equation}\label{EX}
f_0^\var(x,k) =\int dx' dk' \delta_\var (x-x') \delta_\var (k-k') g_0(x',k'),
\end{equation}
where $\delta_\var (x-x') \delta_\var (k-k')$ is the Wigner transform of a coherent state centered in $(x', k') $. (see for instance \cite{PPpaper}). We observe that in this case the exponent $\alpha$ in (\ref{fgclose}) is equal to $1$.

In this case, any of the families $\{ g_0^\var \}_\var$  in Theorem \ref{MAINteoState} will converge to $g_0$ in $L^2(\R^3\times\R^3)$.
\end{rem}

\begin{rem}\label{relative}
Consider the case of a superposition of coherent states discussed in Remark
\ref{ex}.  For a pure coherent state centered in $(x', k') $ and described by a density matrix $A^{\varepsilon}
_{x',k'} $ and a Wigner function $f^{\varepsilon} _{x',k'}(x,k)$, we have
$$
\| A^{\varepsilon} _{x',k'} \|_{HS}=1,\,  \ \ \| f^{\varepsilon} _{x',k'}
\|_{L^2(\R^3\times\R^3)}=\frac {1}{\left(2\pi \varepsilon\right)^{3/2}}.
$$
In contrast,  when we deal with a mixture of coherent states (as in (\ref{EX})), we have
$ \| f^{\varepsilon} _{0} \|_{L^2}\leq C$ ($C$ independent of
$\varepsilon$). As a consequence  (see (\ref{trhssomat})),  the corresponding density matrix
$A^{\varepsilon}$ has vanishing Hilbert-Schmidt norm.
This is the reason why we consider the relative error in (\ref{eq: teo01}) (the situation we take into account is the mixed state one).

\end{rem}


\section{The Hartree dynamics in the Wigner picture}\label{hartreee}

The semiclassical Hartree equation (with unit mass, $m=1$)  for a pure state  is
\begin{equation}\label{eq: 1semi}
i\var\partial_t u^\var = -\frac{\var^2\Delta}{2}u^\var + \left(\phi\star \vert u^\var\vert^2\right)u^\var.
\end{equation}
For a mixed state, we have to pass to the von Neumann equation for the kernel  $\rho^\varepsilon(x,y,t)$,
\begin{equation}
\label{VonNeu}
\begin{array}{c}
i\varepsilon \frac{\partial }{\partial t}\rho^\varepsilon = 
\left[{
 -\frac{ \varepsilon^2}{2}\Delta_x +\left(\phi\star \rho_t^\var\right)(x)
- \left({ -\frac{ \varepsilon^2}{2}\Delta_y + \left(\phi\star \rho_t^\var\right)(y)  }\right) }\right]\rho^\varepsilon, 
\end{array}
\end{equation}
where we put:
\begin{equation}\label{RHO000}
 \rho^\varepsilon_t(z):=\rho^\varepsilon(z,z,t),
\end{equation}
namely  $\rho^\varepsilon_t(z)$ is the spatial probability density associated with the quantum state described by $\rho^\varepsilon(x,y,t)$.

\vskip 0.25 cm
The Wigner Transform of the wavefunction $u^\varepsilon(x,t)$ is defined  by,
\begin{equation}\label{pure}
 W^\var[u^\var](x,k,t):=\left(\frac 1 {2\pi}\right)^3
 \int\limits_{\mathbb{R}^3} { e^{ i yk}\bar{u}^\varepsilon (x+\varepsilon \frac{y}2,t) u^\varepsilon (x-\varepsilon \frac{y}2,t) \ud y }.
\end{equation}
More generally for a mixed state described by a \emph{density matrix}, namely a positive trace class operator $A^\var(t)$ with kernel $\rho^\varepsilon(x, y, t)$,
 the Wigner function $f^\var$  is
\begin{equation}\label{mixed}
 f^\var(x,k,t)=\left(\frac 1 {2\pi}\right)^3 \int\limits_{\mathbb{R}^3} { e^{i yk} \rho^\varepsilon(x+\varepsilon \frac{y}2,x-\varepsilon \frac{y}2,t) \ud y }.
\end{equation}
 The Wigner function is intimately related to the Weyl symbol, but one should be cautious with the scaling in the Planck constant:
\begin{eqnarray}
\label{eqeagmal}
&&(A^\var(t)\,  G)(x)=\int{ f^\var(\frac{x+y}{2},k,t)e^{i(x-y)\frac{k}{\var}}G(y)dkdy} = Op^{Weyl}(\var^{3}f^\var)G(x)\ \ \ \ \text{for}\ G\in L^2(\R^3).\nonumber\\
&&
\end{eqnarray}

The spatial probability density can be easily expressed in terms of the Wigner function as well, namely
\begin{equation}\label{RHO}
 \rho^\varepsilon_t(z):=\rho^\varepsilon(z,z,t)=\int_{\R^3} f^\var(z,k,t)\ud k,\ \ \ \forall\ \ z\in\R^3.
\end{equation}

Defining 
\begin{equation}\label{eq5wwt6}
V(x,t)=\int_{\R^3}\phi(x-x')\rho_t^\var(x')\ud x'=\int_{\R^3}\phi(x-x')\int_{\R^3} f^\var(x',k,t)\ud k\ \ud x',
\end{equation}
the self-consistent Hartree potential,  the Wigner function evolves according to the Wigner equation,
\begin{equation}
\label{WignerEq}
\begin{array}{c}
\partial_t f^\var +  k \cdot \partial_x f^\var =T_\var^f\, f^\var,
\end{array}
\end{equation}
where, for any $w\in L^2(\R^3\times\R^3)$
\begin{eqnarray}\label{Tnew}
\left(T_\var^f w\right)(x,k)=\frac{i}{(2\pi)^3}\int_{-1/2}^{1/2}\ud\lambda\ \int_{\R^3} \ud S\ \widehat{V}(S,t)
\ e^{ i S x} \ \left(S\cdot\pa_k\right) w(x,k+\var\lambda S,t),
\end{eqnarray}
and $\widehat{V}(S,t)$ is the Fourier transform (with respect to the space variable) of the potential $V(x,t)$ defined in (\ref{eq5wwt6}). It will be useful to observe that (\ref{eq5wwt6}) implies
\begin{equation}\label{FourierPOT}
\widehat{V}(S,t)=\widehat{\phi}(S)\,\,\widehat{\rho}_t^\var (S).
\end{equation}
 \\

It is well known (and easy to check), that the dynamics (\ref{WignerEq}) preserves the integral of the Wigner function $f^\var$ on the phase space $\R^3\times\R^3$ i.e. the trace (see \cite{Bove} and Lemma \ref{lml1reg} below). 
This corresponds to the conservation of the  $L^2$-norm of 
the wave function $u^\var$, in case of a pure state, or to the conservation of the trace in case of a density matrix. For this reason we will have
\begin{equation}\label{presINT1}
 \int_{\R^3}\int_{\R^3}f^\var(x,k,t)\ud k \ \ud x=\int{\rho^\var(x,x,t)dx}=1,\ \ \ \text{for any }\ t\geq 0.
\end{equation}

\vskip 0.3cm
\begin{lemma}[$L^1$ regularity] \label{lml1reg} Consider the initial value problem for equation (\ref{WignerEq}) with initial datum $f_0^\var$. Under our assumptions on $\phi$ (more precisely, it is sufficient that $\widehat{\phi} \in L^1(\R^3)$), the trace is preserved by the time evolution. Moreover, if $f_0^\var \in L^1(\R^3\times\R^3)$, $f^\var(t)$ stays in $L^1(\R^3\times\R^3)$ for all $t \in \mathbb{R}$.
\end{lemma}

\begin{dem}
The trace associated with a Wigner function $f^\var(t)$, namely, $I(f^\var(t))=tr \left({Op^{Weyl}(\var^3 f^\var(t))}\right)$, is easily seen to be
\begin{equation}
I(f^\var(t))=\int{ \, \int{f^\var(x,k,t)dk}\, dx}= \int{ \rho^\var_t(x) dx}
\end{equation} 
where the $dx$ integral is understood to be absolutely convergent. The result for the preservation of the trace itself can be found in \cite{Bove} (under the assumption $\phi\in L^\infty$, which obviously holds in this context). So far the $dk$ integral does not have to be absolutely convergent, but only Cauchy-PV.

Now, since $\rho_t^\var \in L^1(\R^3)$ for any $t$ (and in particular $\left\| \rho_t^\var \right\|_{ L^1(\R^3)}=\left\| \rho_0^\var \right\|_{ L^1(\R^3)}=1$), it follows that $\widehat{\rho_t^\var} \in L^\infty(\R^3)$ for any $t$ (indeed $\|\widehat{\rho_t^\var}\|_{L^\infty(\R^3)}\leq 1$),  and therefore 
$$
||\widehat{V}||_{L^1(\R^3)} \leqslant ||\hat{\rho}_{t}^\var||_{L^\infty(\R^3)} \,\,||\widehat{\phi}||_{L^1(\R^3)}\leq ||\widehat{\phi}||_{L^1(\R^3)}.
$$
Now we can rewrite the Wigner equation (\ref{WignerEq}) as 
\begin{equation}
\label{eql1WignerEq}
(\partial_t + k\cdot\partial_x)f^\var=\frac{i}{(2\pi)^3} \int{\widehat{V}(S)e^{iSx} \frac{f^\var(x,k+\frac{\var S}2,t)-f^\var(x,k-\frac{\var S}2,t)}{\var}dS}.
\end{equation}
One readily observes that the $L^1$-norm (with respect to $x$ and $k$) of the rhs of (\ref{eql1WignerEq}) is bounded by
\begin{eqnarray}
\frac{2}{\var(2\pi)^3}||\widehat{\phi}||_{L^1(\R^3)}\,\, ||f^\var||_{L^1(\R^3\times\R^3)}.
\end{eqnarray}
Since the free propagator (associated with the lhs of (\ref{eql1WignerEq})) preserves the $L^1$-norm, the result follows by applying the Gronwall lemma. Observe that the constant grows like $e^{\frac{C}{\var}t}$ (i.e. diverging behaviour as $\var\to 0$) but this does not play any role here and we get all we need in justifying that the phase-space integral of $f^\var$ is absolutely convergent. 
\end{dem}

In the present context we need extra regularity properties in the framework of the Wigner formalism. Actually we can establish the following

\begin{lemma}[Sobolev regularity]\label{lemsrw}
Assume the potential $\phi$ to satisfy the condition
$$
\int d S \vert \widehat{\phi}(S)\vert\, \vert S\vert^{m+1}<+\infty,
$$
for some $m\geq 0$.
Then, for any $T>0$, there is a constant $C$ such that
\begin{equation}\label{sobOK}
||f^\varepsilon(T)||_{H^m(\R^3\times\R^3)} \leqslant ||f^\varepsilon(0)||_{H^m(\R^3\times\R^3)} e^{CT}
\end{equation}
In particular for $m=0$ (i.e. looking at the $L^2$-norm), we have
\begin{equation}
||f^\varepsilon(T)||_{L^2(\R^3\times\R^3)} = ||f^\varepsilon(0)||_{L^2(\R^3\times\R^3)}.
\end{equation}
\end{lemma}

\noindent {\bf Proof:} By using the same observation as before, i.e. that $|\widehat{V}(S)|\leqslant ||\widehat{\rho}_t^\var||_{L^\infty}  |\widehat{\phi}(S)|\leqslant |\widehat{\phi}(S)|$, it follows that equation (\ref{WignerEq}) can be treated in the same way as a problem with a smooth, time-dependent potential. The proof for the corresponding time-independent linear problem \cite{PULVIRENTI,AP} can be adapted to that end in a straightforward manner -- this can be seen in more detail e.g. in \cite{Pezzotti}.

\section{Husimi Transform and Husimi dynamics} \label{secthusimi007}

 Given a Wigner function  $f^\var$  associated to a physical state (pure or mixed),  we define the Husimi transform (HT)  as

\begin{eqnarray}
\label{eqDefPhi}
 \widetilde{f}^\var (x,k)= &&\mathcal{F}^{-1}_{a,b \rightarrow x,k} [ e^{-\frac{1}{4}( \var a^2 + \var b^2 )} \mathcal{F}_{z,y \rightarrow a,b} [f(z,y)]  ]=\nonumber\\
&& = \frac 1{(\pi\var)^3 }
\int_{\R^3\times\R^3}e^{- \frac{(x-x')^2}{\var}- \frac{(k-k')^2}{\var}}f(x',k')\ud x'\ud k'
\end{eqnarray}

We sometimes denote by $\Phi$ the smoothing map
$$
f \to \Phi (f)= \widetilde f.
$$

The remarkable feature of the HT 
relies on the fact that 
$ \widetilde{f}^\var \geq 0$ (see also the discussion in the Introduction). Indeed, in case of a pure state described by a wave function $u$, it can be verified by direct computation that
\begin{eqnarray}
\widetilde{f}^\var(x,k)&&=\frac 1{(\pi\var)^3 }
\int_{\R^3\times\R^3}e^{- \frac{(x-x')^2}{\var}- \frac{(k-k')^2}{\var}}W^\var[u](x',k')\ud x'\ud k'=\nonumber\\
&&=\left\vert\int d x'\, \overline{u}(x') \frac{1}{(\pi\var)^{\frac{3}{4}}}\,   e^{\frac{i}{\var}k'(x-x' )}\, e^{-\frac{(x-x' )^2}{2\var}}\right\vert^2\geqslant 0.
\end{eqnarray}
More generally, 
\begin{equation}
\widetilde{f}^\var(x,k) =\frac{1}{(\pi\var)^{\frac{3}{2}}}\,  
\langle{ A^\var   
e^{\frac{i}{\var}k'(x-x' )}\, e^{-\frac{(x-x' )^2}{2\var}}, e^{\frac{i}{\var}k'(x-x' )}\, e^{-\frac{(x-x' )^2}{2\var}} }\rangle_{L^2_{x'}},
\end{equation} 
where $A^\var$ is the density matrix associated with the Wigner function $f^\var$.
\vskip 0.3cm
Applying the map $\Phi$ to the Wigner equation (\ref{WignerEq}), one finds \cite{AMP}
\begin{equation}
\label{WignerEqSWT}
\partial_t \widetilde{f}^\var + \left({  k \cdot \partial_x +\frac{\var}{2}\, \partial_x\cdot\partial_k }\right) \widetilde{f}^\var=\widetilde{T}_\var^f\, \widetilde{f}^\var,
\end{equation}
where
\begin{eqnarray}\label{Ttilde}
\widetilde{T}_\varepsilon^f \widetilde{w}=\frac{i}{(2\pi)^3}\int_{-1/2}^{1/2}d\lambda\, \int_{\R^3}dS\, \widehat{\widetilde{V}}(S,t)\, e^{iSx}\, (S\cdot \pa_k)\, \widetilde{w}(x+i\frac{\var}{2} S, k+\var\lambda S).
\end{eqnarray}

The key observation is that, {\em up to a small error}, equation (\ref{WignerEqSWT})
can be recasted as
\begin{equation}
\label{WignerEqSWTerror}
\partial_t \widetilde{f}^\var +  k \cdot \partial_x  \widetilde{f}^\var=T_\var^{\widetilde f} \widetilde{f}^\var+ E(t),
\end{equation}
where 
\begin{equation}
\label{error}
E(t)=E_1(t)+E_2(t)
\end{equation}
and
\begin{equation}
\label{error1}
E_1=- 
\frac{\var}{2}\, \partial_x\cdot\partial_k  \widetilde{f}^\var , \qquad  E_2=\widetilde{T}^f_\var \widetilde{f}- T_\var^{\widetilde f} \, \widetilde{f}^\var=\Phi (T^f_\var f)- T_\var^{\widetilde f} \, \widetilde{f}^\var.
\end{equation}
It is straightforward to observe that, according to (\ref{Tnew}) (and (\ref{FourierPOT})), we have
\begin{eqnarray}\label{Tnew2}
\left(T_\var^{\widetilde{f}} \widetilde{w}\right)(x,k)=\frac{i}{(2\pi)^3}\int_{-1/2}^{1/2}\ud\lambda\ \int_{\R^3} \ud S\ \widehat{\widetilde{V}}(S,t)
\ e^{ i S x} \ \left(S\cdot\pa_k\right) \widetilde{w}(x,k+\var\lambda S,t),
\end{eqnarray}
where
\begin{eqnarray}\label{potenzialeS1}
\widehat{\widetilde{V}}(S,t)=e^{-\frac{\var S^2}{4}}\widehat{V}(S,t)=\widehat{\phi}(S)\, \widehat{\widetilde{\rho}}^{\var}_t(S)=\widehat{\phi}(S)\, \widehat{\rho}^{\widetilde{f}}_t(S),
\end{eqnarray}
and the last equality follows easily by direct computation by setting:
\begin{eqnarray}\label{potenzialeS2}
\rho^{\widetilde{f}}_t(x)=\int dk\, \widetilde{f}^\var(x,k,t).
\end{eqnarray}
This observation will lead to the proof of Lemma \ref{mainlemma}.

The point here is that, up to a small error, the (non-negative) Husimi function satisfies the same nonlinear equation with self-consistent potential as the Wigner transform does. Indeed this observation was one of the main findings of \cite{AMP,AP}, namely that (at least formally) to the leading order the Husimi equations are like the Wigner equations, but the potential has been replaced by a mollified version of itself. That is used here, since the $dk$ marginal of the Husimi function is the mollification of the marginal of the Wigner function. Thus we preserve the structure of the quantum phase-space equation, while we change our function with one that remains non-negative.

It must be noted that, once we formulate $E_2=\Phi (T^f_\var f)- T_\var^{\widetilde f} \, \widetilde{f}^\var$, we do not really need the infinite order machinery to proceed (to the proof of Lemma \ref{mainlemma} in this case).

\section{Vlasov evolution}\label{vlasovv}

The Vlasov equation describes the situation in which we have a large number of classical particles in a mean-field regime (see for instance \cite{BH,McK,Sp}, for the case of smooth potentials, and \cite{JABIN}, for more singular interactions).

Denoting by $\Phi^t_V(x,k)$ the flow associated with the system:
\begin{equation}\label{eq: NewtonVlasov}
\left\{
   \begin{aligned}
   & \dot{x} =  k,\\
   & \dot{k}=  -  \pa_{x} \phi\star \rho_t^{g^\var},
   \end{aligned}
   \right.
 \end{equation}  
one can easily verify that the solution $g^\var(t)$ of (\ref{eq: 21}) with initial datum $g_0^\var$ (see the claim of the Theorem \ref{MAINteoState} ) is obtained by propagating the initial datum through the characteristic curves of the flow $\Phi^t_V(x,k)$, namely.
\begin{equation}\label{eq: class6}
g^\var(x,k;t)=g_0^\var\left(\Phi^{-t}_V\left(x,k\right)\right).
\end{equation}
Therefore in proving existence and uniqueness of the solution of (\ref{eq: 21}) one has to deal with a system of ODEs with a self-consistent field (see (\ref{eq: NewtonVlasov})) and the smoothness of the potential $\phi$ is  sufficient to apply a fixed point argument (see \cite{BH, DOB,Neun}) .

Also for the Vlasov equation we need some regularity properties of the solution and we will make use of the following

\begin{lemma}[Vlasov regularity]\label{lemsrv}
Assume the potential  $\phi\in C_b^{m+1}(\R^3)$ for some $m\geq 1$.  Then  there is a constant $C$
such that
\begin{equation}\label{sobOK1}
||g^\var(t)||_{H^m(\R^3\times\R^3)} \leqslant ||g_0^\var||_{H^m(\R^3\times\R^3)} e^{Ct}
\end{equation}
\end{lemma}

The same proof holding for the Wigner case does apply here.

\section{Proof of Theorem \ref{MAINteoState}}
\label{setionproof765}

In the course of the proof we will  denote by $C$ any positive constant,  possibly depending on $\phi$, $f_0^\var$ or $g_0^\var$, but neither on $t$ nor on $\var$.

It is not difficult to show that under suitable smoothness assumptions the Wigner and Husimi functions are close in $L^2$.  In particular, by direct computation the following inequality can be proven

\begin{equation}\label{close}
\left\| \widetilde f^\var (t) -f^\var (t) \right\|_{L^2(\R^3\times\R^3)}\leq C\, e^{C\, t} \var,
\end{equation}
 (see also Lemma A.1 in \cite{AP}).\\

Moreover we can also show that $\widetilde f^\var$ practically solves the nonlinear Wigner equation up to a small error in $L^2$ (see (\ref {WignerEqSWTerror})). This will be used in the main body of the proof below. 

Indeed, we recast equation (\ref{WignerEqSWTerror}) as
\begin{equation}
\label{WignerEqSWTerror007}
\partial_t \widetilde{f}^\var +  k \cdot \partial_x  \widetilde{f}^\var-T_\var^{\widetilde f} \widetilde{f}^\var= E(t),
\end{equation}
(see Section \ref{secthusimi007}) and observe that  (\ref {WignerEqSWTerror007}) can be seen as a Wigner equation with a  time-dependent source term. The error $E(t)$ is ensured to be small by the following:

\begin{lemma} \label{mainlemma} Assume $f_0^\var$ to satisfy all the assumptions of Theorem \ref{MAINteoState},
and $E(t)$ be defined by equations (\ref {WignerEqSWTerror})-(\ref{error1}). Then
\begin{eqnarray}\label{errore}
\left\|E(t)\right\|_{L^2(\R^3\times\R^3)}\leq C\, e^{C\, t}\,  \var.
\end{eqnarray}
\end{lemma}

\begin{dem}
We first bound $E_1$ (see (\ref {error}) and (\ref {error1})).
We have
\begin{eqnarray}
\label{errore2}
\left\|E_1(t)\right\|_{L^2(\R^3\times\R^3)} 
= \frac{\var}{2} \left\|  \pa_x \cdot \pa_k  \widetilde f^\var \right\|_{L^2(\R^3\times\R^3)}
\leq \frac{\var}{2}  \left\|  f^\var \right\|_{H^2(\R^3\times\R^3)}
\leq C\, e^{C\, t} \var,
\end{eqnarray}
where we 
estimated, uniformly in $\var$, the $H^2$-norm of $\widetilde f^\var $ with the $H^2$-norm of $f^\var $ and we used property (\ref{sobOK}) for $f^\var$. 

Moreover, since we have 
\begin{eqnarray}\label{E2del}
E_2= \widetilde{T}^f_\var\,  \widetilde{f}^\var- T_\var^{\widetilde f} \, \widetilde{f}^\var,
\end{eqnarray}
then, by (\ref{Tnew2}) and (\ref{Ttilde}) we get
\begin{eqnarray}
\label{difference}
E_2=\frac{i}{(2\pi)^3}\int_{-1/2}^{1/2}d\lambda\, \int_{\R^3}dS\, \widehat{\widetilde{V}}(S,t)\, e^{iSx}\, (S\cdot \pa_k)\, \left[\widetilde{f}^\var(x+i \frac{\var}{2}S, k+\var\lambda S)-\widetilde{f}^\var(x, k+\var\lambda S)\right]
\end{eqnarray}
and we remind that $E_2=E_2(x,k,t)$. Thus, by taking the Fourier transform $\mathcal{F}_{x,k\to p,q}$ 
we find:
\begin{eqnarray}
\label{T7}
\widehat{E_2} (p,q)= \frac{i}{(2\pi)^3}\int_{-1/2}^{1/2} d\lambda\, \int_{\R^3} dS\,  \widehat{\phi}(S)\, \widehat{\rho}_t^{\widetilde{f}}(S)\,  e^{i  \var \lambda  Sq} \, S\cdot (iq)\, \widehat{\widetilde{f}^\var} ( p-S,q)
\left[ e^{\frac{\var}{2}\,  S^2} e^{-\frac{\var}{2} pS} -1\right],
\end{eqnarray}
namely
\begin{eqnarray}
\label{T7bis}
\widehat{E_2} (p,q)= \frac{i}{(2\pi)^3} \int_{-1/2}^{1/2} d\lambda\, \int_{\R^3} dS\,  \widehat{\phi}(S)\, \widehat{\rho}_t^{f}(S)\,  e^{i  \var \lambda  Sq} \, S\cdot (iq)\, \widehat{f}^\var ( p-S,q)e^{-\frac{\var}{4} p^2} e^{-\frac{\var}{4} q^2}
\left[ 1-e^{-\frac{\var}{2}\,  S^2} e^{ \frac{\var}{2}pS} \right],
\end{eqnarray}

By applying the Taylor formula, for some $\xi \in (0,\var)$ we get
\begin{eqnarray}
\label{Tay}
\left[1 -e^{- \frac{\var}{2}S^2} e^{\frac{\var}{2} pS}  \right] = \frac{\var}{2}  (S^2-p \cdot S) e^{-\frac{\xi}{2}S^2} e^{\frac{\xi}{2}  pS} 
\end{eqnarray}
and hence
\begin{eqnarray}
\label{Tay2}
\left|e^{-\frac{\var}{4} p^2} e^{-\frac{\var}{4} q^2} \left[ 1- e^{-\frac{\var}{2}  S^2} e^{ \frac{\var}{2} pS}  \right] \right| \leq \frac{\var}{2}  (S^2+|p| |S|) \leq  \var (S^2+|p-S| |S|) .
\end{eqnarray}
Finally we obtain
\begin{eqnarray}
\label{est1}
|\widehat{E_2}  (p,q)|^2 \leq C \var^{2}\, \vert q\vert^2
\left (  \int dS |\widehat \phi(S)|\, \vert S\vert \, ( S^2+|p-S|\vert S \vert )\,  |\widehat{f}^\var ( p-S,q)| 
\right )^2,
\end{eqnarray}
where we used (\ref{Tay2}) and the uniform $L^\infty$ control on $\widehat{\rho}_t^{f}$.

Then, by applying the Cauchy-Schwarz inequality
\begin{eqnarray}
\label{est2}
&&| \widehat{E_2}  (p,q)|^2 \leq C \var^{2}\, \vert q\vert^2
  \left(\int dS |\widehat \phi(S)|\vert S\vert^2 \right)\left(\int dS |\widehat \phi (S)| ( S^2+|p-S|\vert S \vert )^2 |\widehat{f}^\var ( p-S,q)|^2\right) .\nonumber\\
&&
\end{eqnarray}
Therefore 
\begin{eqnarray}
\label{est3}
\left\|E_2\right\|_{L^2(\R^3\times\R^3)} \leq C \var\, \left(\int dS |\widehat \phi (S)| |S|^2 \right)
\left(\int dS |\widehat \phi (S)| |S|^4 \right)\left\| f^\var \right\|_{H^2(\R^3\times\R^3)}
\leq C\, e^{C\, t} \var,\nonumber\\
&&
\end{eqnarray}
where we made use of property (\ref{sobOK}) for $f^\var$ and the assumptions we did on $\phi$ (see (\ref{MOM4})). \\

Then, by (\ref{errore2}) and (\ref{est3}) we obtain that (\ref{errore}) holds true and the proof of Lemma \ref{mainlemma} is concluded. 
\end{dem}

\vskip 0.2cm
Before moving on, let us make a remark on notation: as it has been explained previously (see e.g. the statement of Theorem \ref{MAINteoState}, or the discussion in Section \ref{sectassumptions00}) the classical initial data $g^\var_0$ may or not depend on $\var$. However all the characteristics of interest here (Sobolev norms, support etc) are bounded uniformly in $\var$. Consistently with that, and for simplicity in the notation, we will drop the superscript $\var$, and refer to the initial data $g_0$ and the classical solution $g(t)$ in the sequel.

For convenience of the reader, we recall that, according to the statement of Theorem \ref{MAINteoState}, $g_0$ satisfies the property:
\begin{equation}\label{gcomp}
supp \, g_0 \subseteq \{ (x,k)\in \R^3\times \R^3 : \, \, \,  |k|\leq M_0 \},
\end{equation}
for a certain constant $M_0$ not depending on $\var$.

\vskip 0.75cm
By virtue of (\ref{close}), in order to prove Theorem \ref{MAINteoState} (more precisely, formula (\ref{MainWIGNER}) ) it suffices to bound in $L^2$ the remainder $h_t(x,k)=\widetilde{f}^\var(x,k,t)-g(x,k,t)$; indeed we will show that
\begin{equation}\label{h}
||h_t||_{L^2(\R^3\times\R^3)} = ||\widetilde{f}^\var(t) - g(t)||_{L^2(\R^3\times\R^3)} \leqslant C e^{ C e^{C t}} \var^{\frac 27 \al}.
\end{equation}
Then, (\ref{MainWIGNER}) will follow straightforward by joining (\ref{close}) and (\ref{h}).

\vskip 0.75cm
\noindent {\bf Proof of equation (\ref{h})}
By (\ref{eq: 21}) and (\ref{WignerEqSWTerror}) the evolution of $h_t$ is given by
\begin{equation}
\label{eq: 24}
\left\{ {
\begin{array}{l}
\pa_t h_t\ + k\cdot \partial_x h_t = T_\var^{\widetilde{f} } h_t+ \left(T_\var^{\widetilde{f} }-T_0^{\widetilde{f} }\right) g+\left(T_0^{\widetilde{f} }-T_0^g\right)g+ E(t) .\\ { } 
h_0(x,k)=\widetilde{f}^\varepsilon_0(x,k)-g_0(x,k).
\end{array}
} \right.
\end{equation}

Let $\Omega^t$  be the ($L^2$ preserving) Wigner-Liouville flow associated with the equation:
\begin{eqnarray}\label{eq: 24bis}
\pa_t h_t\ +  k\cdot \partial_x h_t = T_\var^{\widetilde{f} } h_t.
\end{eqnarray}
Then, from (\ref{eq: 24}) 
we have:
\begin{eqnarray}\label{eq: 25}
 h_t=\Omega^{t} h_0 + \int_0^t \ud s\ \Omega^{t-s}(E(s)) + \int_0^t \ud s\ \Omega^{t-s} \left(r_1(s) + r_2(s))\right),
\end{eqnarray}
where (see  (\ref{Tnew}) and (\ref{eq: 22}))
\begin{eqnarray}\label{eq: 26}
 r_1(s)&:=&\left(T_\var^{\widetilde{f} }-T_0^{\widetilde{f} }\right) g(x,k,s) =\nonumber\\
&& = \frac{i}{(2\pi)^3}\int_{-1/2}^{1/2}\ud \lambda\ \int_{\R^3} \ud S\ \widehat{\phi}(S)\widehat{\rho}_s^{\widetilde{f} } (S)\ e^{i S x} \ \left(S\cdot\partial_k\right)\left[g(x,k+\var\lambda \ S,s)-g(x,k,s)\right],\nonumber\\
&&
\end{eqnarray}
and
\begin{eqnarray}\label{eq: 27}
&& r_2(s):=\left(T_0^{\widetilde{f} }-T_0^g\right) g(x,k,s) = \left(\int_{\R^3} \ud y\ \partial_x\phi(x-y)\left[\rho_s^{\widetilde{f} }(y)- \rho_s^g(y)\right]\right)\cdot \partial_k g(x,k,s).\nonumber\\
&&
\end{eqnarray}

Next, from (\ref{eq: 25}), we have:
\begin{eqnarray}\label{eq: 28}
 \left\|h_t\right\|_{L^2(\R^3\times\R^3)}&\leq &\left\|h_0 \right\|_{L^2(\R^3\times\R^3)} +\int_0^t \ud s\ \left\|E(s)\right\|_{L^2(\R^3\times\R^3)}+\nonumber\\
 &&+\int_0^t \ud s\ \left(\left\|r_1(s)\right\|_{L^2(\R^3\times\R^3)} + \left\|r_2(s)\right\|_{L^2(\R^3\times\R^3)}\right),
\end{eqnarray}
and, by Lemma \ref{mainlemma} 
\begin{eqnarray}\label{eq: 28bis}
 \left\|h_t\right\|_{L^2(\R^3\times\R^3)}\leq \left\|h_0 \right\|_{L^2(\R^3\times\R^3)} + C(e^{C\, t}+1)\,  \var +\int_0^t \ud s\ \left(\left\|r_1(s)\right\|_{L^2(\R^3\times\R^3)} + \left\|r_2(s)\right\|_{L^2(\R^3\times\R^3)}\right).\nonumber\\
\end{eqnarray}


Moreover by  (\ref{fgclose}) and  estimate (\ref{close}), we easily get:
\begin{eqnarray}\label{eq: 28new}
&&\left\|h_0 \right\|_{L^2(\R^3\times\R^3)} = \left\|\widetilde{f}_0^\var-g_0\right\|_{L^2(\R^3\times\R^3)} \leq \left\|\widetilde{f}_0^\var-f_0^\var\right\|_{L^2(\R^3\times\R^3)}+\left\|f_0^\var-g_0\right\|_{L^2(\R^3\times\R^3)}= C\, \var^{\alpha}.\nonumber\\
&&
\end{eqnarray}

 Then, we finally get:
 
 \begin{eqnarray}\label{eq: 28bisII}
 \left\|h_t\right\|_{L^2(\R^3\times\R^3)}\leq C(e^{C\, t}+1) \var^{\alpha} +\int_0^t \ud s\ \left(\left\|r_1(s)\right\|_{L^2(\R^3\times\R^3)} + \left\|r_2(s)\right\|_{L^2(\R^3\times\R^3)}\right).
\end{eqnarray}

Next we evaluate  the $L^2$-norm of $r_1(s)$. 

We observe that, by virtue of the positivity of $\widetilde{f}^\var$, the $L^\infty$-norm of $\widehat{\rho}_t^{\widetilde{f}}$ is uniformly bounded. In fact, we have:
\begin{eqnarray}\label{eq: 28newI}
\left\|\widehat{\rho}_t^{\widetilde{f}}\right\|_{L^\infty(\R^3)}\leq \left\|\rho_t^{\widetilde{f}}\right\|_{L^1(\R^3)}=\int_{\R^3}\ud x\ \left\vert\int_{\R^3} \ud k\, \widetilde{f}^\var(x,k,t)\right\vert=1,
\end{eqnarray}
where the last equality is easily obtained by direct computation (see (\ref{eqDefPhi}) and (\ref{presINT1}) ).
Then, by applying the Taylor formula in (\ref{eq: 26}), we can estimate the $L^2$-norm of $r_1(s)$ as follows:
\begin{eqnarray}\label{eq: 29}
&& \left\|r_1(s)\right\|_{L^2(\R^3\times\R^3)}\leq C \var^2 \left(\int_{\R^3} \ud S\, \vert \widehat{\phi}(S)\vert\, \vert S\vert^3\right)\, \left\| g(s)\right\|_{H^3(\R^3\times\R^3)}\leq C\, e^{C\, s}\var^2,
\end{eqnarray}
where, in the last inequality, we used property (\ref{sobOK1}) for $g(s)$ and assumption (\ref{MOM4}) for $\phi$.\\
Then, by (\ref{eq: 28bisII}) we get
\begin{eqnarray}\label{eq: 28bisIII}
 \left\|h_t\right\|_{L^2(\R^3\times\R^3)}\leq C(e^{C\, t}+1) \var^{\alpha} +C(e^{C\, t}+1) \var^{2} +\int_0^t ds\,  \left\|r_2(s)\right\|_{L^2(\R^3\times\R^3)}.
\end{eqnarray}

Now, let us look at the $L^2$-norm of $r_2(s)$. 

By (\ref{eq: 27}) we have
\begin{eqnarray}\label{eq: 32}
 &&\left\|r_2(s)\right\|_{L^2(\R^3\times\R^3)}^2 \leq \nonumber\\
 &&\leq\int_{\R^3} \ud x\int_{\R^3} \ud k\left\vert\partial_k g(x,k,s)\right\vert^2\left(\int_{\R^3} \ud y\ \left\vert\partial_x\phi(x-y)\right\vert \int_{\R^3} \ud w \left\vert \widetilde{f}^\var(y,w,s)-g(y,w,s)\right\vert\right)^2.\nonumber\\
&&
\end{eqnarray}
We split the integral
\begin{eqnarray}\label{eq: 33}
&&   \int_{\R^3} \ud w \left\vert \widetilde{f}_s(y,w)-g_s(y,w)\right\vert
\end{eqnarray}
into the two domains $\vert w\vert\leq M$ and $\vert w\vert>M$,  where $M$ is chosen in the following way. \\If $X(t),K(t) $  is the classical flow generated by the force field
$$
-\int dy \partial_x\phi(x-y)\zeta (y,t),
$$
where $\zeta$ is any spatial probability density, then
$$
|K(t)| \leq |K(0)|+ \| \partial_x\phi \|_{L^\infty} t.
$$
Therefore,  by virtue of assumption (\ref{gcomp}) on the initial datum $g_0$, there exists a positive constant $M$ for which (for $t\leq T$ arbitrary but fixed)
\begin{equation}
\label{support}
g(x,k,t)=0  \qquad \hbox{if} \quad  |k|>M.
\end{equation}
Clearly $M=M(t)$ depends on time and it is straightforward to check that:
\begin{equation}
\label{M}
M(t)=M_0+ \left\|\pa_x\phi\right\|_{L^\infty} t,
\end{equation}
where $M_0$ is the same as in (\ref{gcomp}).

Hence, by (\ref{eq: 32}),
we have:
\begin{eqnarray}\label{eq: 34ante}
&&\left\|r_2(s)\right\|_{L^2(\R^3\times\R^3)}^2\leq \nonumber\\
&&\leq \int_{\R^3} \ud x\int_{\R^3} \ud k \left\vert\partial_k g(x,k,s)\right\vert^2\left(\int_{\R^3} \ud y\ \left\vert\partial_x\phi(x-y)\right\vert \int_{\vert w\vert\leq M} \ud w \left\vert \widetilde{f}^\var(y,w,s)-g(y,w,s)\right\vert\right)^2+ \nonumber\\
&&+\int_{\R^3} \ud x\int_{\R^3} \ud k \left\vert\partial_k g(x,k,s)\right\vert^2\left(\int_{\R^3} \ud y\ \left\vert\partial_x\phi(x-y)\right\vert \int_{\vert w\vert> M} \ud w \left\vert \widetilde{f}^\var(y,w,s)-g(y,w,s)\right\vert\right)^2+\nonumber\\
&&+2\int_{\R^3} \ud x\int_{\R^3} \ud k \left\vert\partial_k g(x,k,s)\right\vert^2\left(\int_{\R^3} \ud y\ \left\vert\partial_x\phi(x-y)\right\vert \int_{\vert w\vert\leq  M} \ud w \left\vert \widetilde{f}^\var(y,w,s)-g(y,w,s)\right\vert\right)\times\nonumber\\
&&\ \times\left(\int_{\R^3} \ud y\ \left\vert\partial_x\phi(x-y)\right\vert \int_{\vert w\vert> M} \ud w \left\vert \widetilde{f}^\var(y,w,s)-g(y,w,s)\right\vert\right) \leq \nonumber\\
&& 2\int_{\R^3} \ud x\int_{\R^3} \ud k \left\vert\partial_k g(x,k,s)\right\vert^2\left(\int_{\R^3} \ud y\ \left\vert\partial_x\phi(x-y)\right\vert \int_{\vert w\vert\leq M} \ud w \left\vert \widetilde{f}^\var(y,w,s)-g(y,w,s)\right\vert\right)^2+\nonumber\\
&&+2\int_{\R^3} \ud x\int_{\R^3} \ud k \left\vert\partial_k g(x,k,s)\right\vert^2\left(\int_{\R^3} \ud y\ \left\vert\partial_x\phi(x-y)\right\vert \int_{\vert w\vert> M} \ud w \left\vert \widetilde{f}^\var(y,w,s)-g(y,w,s)\right\vert\right)^2.\nonumber\\
&&
\end{eqnarray}

We first bound the first term on the right hand side of (\ref{eq: 34ante}).
We find
\begin{eqnarray}\label{eq: 36}
&&   \left(\int_{\R^3} \ud y\ \left\vert\partial_x\phi(x-y)\right\vert\int_{\vert w\vert\leq M} \ud w \left\vert \widetilde{f}^\var(y,w,s)-g(y,w,s)\right\vert\right)^2=\nonumber\\
&&\ = \left(\int_{\R^3} \ud y\int_{\R^3} \ud w \ \left\vert\partial_x\phi(x-y)\right\vert\ \chi_{\vert w\vert\leq M}(w)\ \left\vert \widetilde{f}^\var(y,w,s)-g(y,w,s)\right\vert\right)^2\leq\nonumber\\
&&\nonumber\\
&&\leq \frac{4}{3}\pi M^{3}\left\vert|\partial_x\phi\right\vert|_{L^2}^2 \ 
\left\|\widetilde{f}^\var(s)-g(s)\right\|_{L^2(\R^3\times\R^3)}^2\leq \frac{4}{3}\pi M^{3}\left\|\partial_x\phi\right\|_{L^2}^2 \ 
\left\|h_s\right\|_{L^2(\R^3\times\R^3)}^2.
\end{eqnarray}

Therefore, 
we obtain
\begin{eqnarray}\label{eq: 37}
&& 2\int_{\R^3} \ud x\int_{\R^3} \ud k \left\vert\partial_k g(x,k,s)\right\vert^2\left(\int_{\R^3} \ud y\ \left\vert\partial_x\phi(x-y)\right\vert \int_{\vert w\vert\leq M} \ud w \left\vert \widetilde{f}^\var(y,w,s)-g(y,w,s)\right\vert\right)^2\leq \nonumber\\
&&\leq Ce^{Cs}\, \ M^{3}\left\|h_s\right\|_{L^2(\R^3\times\R^3)}^2= C e^{C \, s}(C+s)^3
\left\|h_s\right\|_{L^2(\R^3\times\R^3)}^2,
\end{eqnarray}
by virtue of
Lemma \ref{lemsrv}, (\ref{M}) and assumption (\ref{phiH1}) on $\phi$.

Now let us look at
\begin{eqnarray}\label{eq: 38}
&&  2\int_{\R^3} \ud x\int_{\R^3} \ud k \left\vert\partial_k g(x,k,s)\right\vert^2\left( \int_{\R^3} \ud y\ \left\vert\partial_x\phi(x-y)\right\vert\int_{\vert w\vert> M} \ud w \left\vert \widetilde{f}^\var(y,w,s)-g(y,w,s)\right\vert\right)^2.\nonumber\\
&&
\end{eqnarray}

By (\ref{support}) we  have
\begin{eqnarray}\label{eq: 39}
&& 2\int \ud x\  \ud k \left\vert\partial_k g(x,k,s)\right\vert^2 \left( \int_{\R^3} \ud y\ \left\vert\partial_x\phi(x-y)\right\vert\int_{\vert w\vert> M} \ud w \left\vert \widetilde{f}^\var(y,w,s)-g(y,w,s)\right\vert\right)^2=\nonumber\\
&& 2\int \ud x\  \ud k \left\vert\partial_k g(x,k,s)\right\vert^2 \left( \int_{\R^3} \ud y\ \left\vert\partial_x\phi(x-y)\right\vert\int_{\vert w\vert> M} \ud w \left\vert \widetilde{f}^\var(y,w,s)\right\vert\right)^2\leq\nonumber\\
&& \leq 4\int \ud x\  \ud k\left\vert\partial_k g(x,k,s)\right\vert^2\left( \int_{\R^3} \ud y\ \left\vert\partial_x\phi(x-y)\right\vert\int_{\vert w\vert\geq \var^{-\ome}} \ud w \left\vert \widetilde{f}^\var(y,w,s)\right\vert\right)^2+\nonumber\\
&& + 4 \int \ud x\  \ud k\left\vert\partial_k g(x,k,s)\right\vert^2\left( \int_{\R^3} \ud y\ \left\vert\partial_x\phi(x-y)\right\vert\int_{M< \vert w\vert\leq \var^{-\ome}} \ud w \left\vert \widetilde{f}^\var(y,w,s)-\overline{g}^\var (y,w,s) \right\vert\right)^2,\nonumber\\
&&
\end{eqnarray}
where $\omega>0$ will be fixed later (see (\ref{eq:omega}))). 
Here $\overline{g}^\var (y,w,s) $ denotes the solution of the classical Liouville equation generated by the force field
$$
-\int  dy  \partial_x\phi(x-y)\rho^{\widetilde f} (y,t),
$$
namely
\begin{eqnarray}\label{vlasov 2}
\pa_t \overline{g}^\var \ + k\cdot \pa_x \overline{g}^\var  = T_0^{\widetilde f} \overline{g}^\var,
\end{eqnarray}
and the initial datum is the same of the Vlasov evolution, namely, $\overline{g}^\var (y,w,0)=g_0(y,w)$.
Note that $\overline{g}^\var$ enters freely in the game because it satisfies  the support property (\ref{support}). \\

Let us estimate the term:
\begin{eqnarray}\label{eq: PREenergy cons}
&&  4\int \ud x\  \ud k\left\vert\partial_k g(x,k,s)\right\vert^2\left( \int_{\R^3} \ud y\ \left\vert\partial_x\phi(x-y)\right\vert\int_{\vert w\vert\geq \var^{-\ome}} \ud w \left\vert \widetilde{f}^\var(y,w,s)\right\vert\right)^2.
\end{eqnarray}
Note that here the positivity of $\widetilde f^\var$ is crucial because it allows us
to use the energy conservation. Indeed
\begin{eqnarray}\label{eq: energy cons}
&& 4\int_{\R^3} \ud x\ \ud k \left\vert\partial_k g(x,k,s)\right\vert^2\left(  \int_{\R^3} \ud y\ \left\vert\partial_x\phi(x-y)\right\vert\int_{\vert w\vert\geq \var^{-\ome}} \ud w \left\vert \widetilde{f}^\var(y,w,s)\right\vert\right)^2\leq \nonumber\\
&& \leq 4\left\|g(s) \right\|_{H^1}^2 \left\|\pa_x\phi\right\|_{L^\infty}^2 \left(\int_{\R^3}  d y \int_{\vert w\vert\geq \var^{-\ome}}  d w\,    \widetilde{f}^\var(y,w,s)\right)^2
\leq \nonumber\\
&&\leq C \var^{4\ome} \left( \int_{\R^3}  d y \int_{\R^3}  d w\,   \frac{w^2}{2}\,   \widetilde{f}^\var(y,w,s)\right)^2\leq C \var^{4\ome}.
\end{eqnarray}
To show the last inequality, we denote  by $\gamma (w,s)=\int dy\,   f^\var(y,w,s)$ the distribution of momenta (which is obviously positive and with integral in $d w$ equal to one) and by $\widetilde{\gamma}(w,s)$ the smoothed version of $\gamma (w,s)$. Then, denoting by $\Phi'$ the smoothing acting only on the momentum variable, we set $\widetilde{\gamma}=\Phi'(\gamma)$ (clearly, the action is exactly as in (\ref{eqDefPhi}) for $\Phi$). As we observed for the spatial distribution $\rho^\var_t$ (see (\ref{potenzialeS1})), even in this case it is straightforward to see that the smoothing commutes with the partial integration on the phase-space, namely
$$
\widetilde{\gamma}(w,s)=\Phi'(\gamma) (w,s)=\Phi'(\int dy\,   f^\var(y,w,s))=\int dy\,  \Phi(f^\var)(y,w,s)=\int dy \widetilde{f}^\var(y,w,s).
$$
Therefore, we can write the term $\int_{\R^3}  d y\int_{\R^3}  d w\,   \frac{w^2}{2}\,   \widetilde{f}^\var(y,w,s)$ in (\ref{eq: energy cons}) as
$$
\int_{\R^3}  d y\int_{\R^3}  d w\,   \frac{w^2}{2}\,   \widetilde{f}^\var(y,w,s)=\int_{\R^3}  d w\,   \frac{w^2}{2}\, \widetilde{\gamma}(w,s).
$$ 
Moreover, we have
\begin{equation}\label{heat}
\int_{\R^3}  d w\,   \frac{w^2}{2}\, \widetilde{\gamma}(w,s)=\int dw \, \frac{w^2}{2}\,  e^{\var\Delta} \gamma (w,s),
\end{equation}
because the action of the smoothing operator is exactly the same of the heat flow $e^{\var\Delta}$ (see (\ref{eqDefPhi})). Now we use the well-known (and easy to check) property
$$
\int dw\,  \frac{w^2}{2}\, e^{\var\Delta} \gamma (w,s)\leq \int dw \left( \frac{w^2}{2}+C\var\right) \gamma (w,s)=\int dw\, \frac{w^2}{2}\, \gamma (w,s) + C\var,
$$
to conclude that:
\begin{equation}\label{heat2}
\int_{\R^3}  d y\int_{\R^3}  d w\,   \frac{w^2}{2}\, \widetilde{f}^\var(y,w,s)\leq \int dw\, \frac{w^2}{2}\, \gamma (w,s) + C\var=\int_{\R^3}  d y\int_{\R^3} dw\,   \frac{w^2}{2}\, f^\var(y, w,s) + C\var.
\end{equation}
Now, 
by the energy conservation:
\begin{eqnarray}\label{eq: energy cons1}
&&\int_{\R^3}  d y \int_{\R^3}   d w\,   \frac{w^2}{2}\,   f^\var(y,w,s)\leq \nonumber\\
&&\leq\int_{\R^3}  d y \int_{\R^3}   d w\,   f^\var(y,w,s)\, \left( \frac{w^2}{2} + \frac{\phi *\rho^f (s)}{2}\right)+C\nonumber\\
&&=\int_{\R^3}  d y \int_{\R^3}   d w\,   f_0^\var(y,w)\, \left( \frac{w^2}{2} + \frac{\phi *\rho^f (0)}{2}\right)+C \leq C,
\end{eqnarray}
where, in the first inequality, we used that the potential $\phi$ is bounded from below ( as a consequence of assumption (\ref{MOM4})) and the last bound follows from our assumptions .

Thus, by (\ref{heat2}) and (\ref{eq: energy cons1}) we get (\ref{eq: energy cons}) (the $L^\infty$ control on $\pa_x\phi$ is guaranteed by (\ref{MOM4}) and the $H^1$ control on $g$ is guaranteed by (\ref{sobOK1})).

Next we estimate the term on the last line of (\ref{eq: 39}), namely:
\begin{eqnarray}\label{eq:40}
&&  4 \int \ud x\  \ud k\left\vert\partial_k g(x,k,s)\right\vert^2 \left(\int_{\R^3} \ud y\ \left\vert\partial_x\phi(x-y)\right\vert\int_{ M< \vert w\vert\leq \var^{-\ome}} \ud w \left\vert \widetilde{f}^\var(y,w,s)-\overline{g}^\var (y,w,s) \right\vert\right)^2 \leq \nonumber \\
&& \leq Ce^{Cs} \var ^{-3 \ome} \left\|\widetilde{f}^\var-\overline{g}^\var\right\|_{L^2(\R^3\times\R^3)}^2.
\end{eqnarray}

To control the $L^2$-norm of  $\widetilde{f}^\var(s)-\overline{g}^\var(s)  $ we set $p^\var_s=\widetilde{f}^\var(s)-\overline{g}^\var(s)$.
Then, by (\ref{WignerEqSWTerror}) and (\ref{vlasov 2}), the equation for $p^\var_s $ is:

\begin{equation}
\label{eq: p}
\left\{ {
\begin{array}{l}
\pa_s p^\var_s\ + k\cdot \partial_x p^\var_s = T_\var^{\widetilde{f} } p^\var_s+ \left(T_0^{\widetilde{f} }-T_\var^{\widetilde{f} }\right) \overline{g}^\var(s)+ E(s) .\\ { } 
p_0^\var(x,k)=\widetilde{f}^\varepsilon_0(x,k)-g_0(x,k).
\end{array}
} \right.
\end{equation}
We proceed as before using that the flow generated by $- k\cdot \partial_x  + T_\var^{\widetilde{f} } $ is isometric in $L^2$.  Therefore 

\begin{equation}
\label{eq: 2p}
\| p^\var_s \|_{L^2} \leq \|p^\var_0 \|_{L^2} + \int _0^s d\tau \| E (\tau)  \|_{L^2}+ \int _0^s 
\|\left(T_0^{\widetilde{f} }-T_\var^{\widetilde{f} }\right) \overline{g}^\var(\tau) \|_{L^2}\,  d\tau
\end{equation}
The first two terms on the right hand side of (\ref{eq: 2p})  have been estimated previously (see (\ref{eq: 28new}) and Lemma \ref{mainlemma}) and they give rise to 
\begin{equation}
\label{eq: 3p}
 \|p^\var_0 \|_{L^2}\leq C\var^{\alpha} \ \ \  \text{and}\ \ \ \int _0^s d\tau \| E (\tau)  \|_{L^2}\leq C(e^{C\, s}+1)\var.
\end{equation}
 Moreover, the last term in (\ref{eq: 2p}) can be estimated exactly as the term $r_1(s)$ (see (\ref{eq: 26}) and (\ref{eq: 29})) because the Liouville dynamics for $\overline{g}^\var$ controls the $H^3$-norm (that is finite and uniformly bounded in $\var$ at time $\tau=0$ since $\overline{g}^\var(0)=g_0$). Therefore we find:
\begin{equation}
\label{eq: 4p}
\int _0^s 
\|\left(T_0^{\widetilde{f} }-T_\var^{\widetilde{f} }\right) \overline{g}^\var(\tau) \|_{L^2}\, d\tau \leq C(e^{C\, s}+1)\var^2.
 \end{equation}
Thus, finally we get:
\begin{equation}
\label{eq: 5p}
\| p^\var_s \|_{L^2} \leq  C(e^{C\, s}+1)\var^{\alpha}.
\end{equation}
Therefore, by setting
\begin{eqnarray}\label{eq:omega}
\omega<\frac{\alpha}{3} 
\end{eqnarray}
we find
\begin{eqnarray}\label{eq:40bisI}
&&  4 \int \ud x\  \ud k\left\vert\partial_k g(x,k,s)\right\vert^2  \left(\int_{\R^3} \ud y\ \left\vert\partial_x\phi(x-y)\right\vert\int_{ M< \vert w\vert\leq \var^{-\ome}} \ud w \left\vert \widetilde{f}^\var(y,w,s)-\overline{g}^\var (y,w,s) \right\vert\right)^2 \leq \nonumber\\
&&\leq C(e^{C\, s}+1)\var^{-3 \ome+\alpha},
\end{eqnarray}
with $-3 \ome+\alpha>0$.
\vspace{0.2cm}\\
In the end, by collecting (\ref{eq: 37}), (\ref{eq: energy cons}) and (\ref{eq:40bisI}), we obtain that:
\begin{eqnarray}\label{eq: 51}
&&  \left\|r_2(s)\right\|_{L^2(\R^3\times\R^3)}^2 \leq C\left(  e^{C\, s}\, (C+s)^3\left\|h_s\right\|_{L^2(\R^3\times\R^3)}^2 + 
\var^{4\omega}+(e^{C\, s}+1)\var^{-3 \ome+\alpha}\right),
\end{eqnarray}
that gives an optimal bound for $\omega=\frac{\alpha}{7}$, namely:
\begin{eqnarray}\label{eq: 52}
&&  \left\|r_2(s)\right\|_{L^2(\R^3\times\R^3)} \leq C\left( e^{C\, s}\, (C+s)^{3/2}\left\|h_s\right\|_{L^2(\R^3\times\R^3)} + 
(e^{C\, s}+1)\var^{\frac{2}{7}\alpha}\right),
\end{eqnarray}
and
\begin{eqnarray}\label{eq: 52bis}
&& \int_0^t ds\,  \left\|r_2(s)\right\|_{L^2(\R^3\times\R^3)} \leq C\left((e^{C\, t}+1+t)\var^{\frac{2}{7}\alpha} + \int_0^t ds\, e^{C\, s}\, (C+s)^{3/2}\left\|h_s\right\|_{L^2(\R^3\times\R^3)} 
\right).\nonumber\\
&&
\end{eqnarray}
\vspace{0.2cm}\\

By (\ref{eq: 28bisIII}) and (\ref{eq: 52bis}), we can finally control the $L^2$-norm of $h_t$, obtaining:
\begin{eqnarray}\label{eq: 53}
 \left\|h_t\right\|_{L^2(\R^3\times\R^3)}
 \leq C \left( (e^{C\, t}+1)\var^{\alpha} + (e^{C\, t}+1+t)\var^{\frac{2}{7}\alpha}+\int_0^t \ud s\ (C+s)^{3/2}\, e^{C\, s} \left\|h_s\right\|_{L^2(\R^3\times\R^3)}\right).\nonumber\\
 &&
\end{eqnarray}
The, by applying the Gronwall lemma, we find that 
\begin{eqnarray}\label{eq: 54}
 && \left\|h_t\right\|_{L^2(\R^3\times\R^3)}\leq C(e^{C\, t}+1+t) \, \var^{\frac{2}{7}\alpha}e^{\int_0^t \ud s\ (C+s)^{3/2}\, e^{C\, s}},
  \end{eqnarray}
 and hence (\ref {h}) holds true.

 Finally, by virtue of (\ref{close}) and (\ref{eq: 54}),  we deduce (\ref{MainWIGNER}) and hence the proof of Theorem \ref{MAINteoState} is concluded.

\qed

\begin{rem}\label{example}
We observe that, for example, in the case in which the initial datum is given by (\ref{EX}), the rate of convergence is
estimated by  $\var^{\frac{2}{7}}$.
\end{rem}

\vspace{0.5cm}

\begin{appendix}
\section{Existence of semiclassical initial data: proof of Proposition \ref{const}} \label{sectappend}

Let $g^\var_0$ be defined by (\ref{ansatz}). Set $N=N(\var):=(\int  \widetilde
f^{\varepsilon}_0(x,k)\chi_{M_0}(|k|)dxdk)^{-1}$. 

Property (\ref{enumitemappend1}) follows by the 
positivity of the Husimi function 
and (\ref{enumitemappend3}) follows by construction.  Since

\begin{equation} \label{eqappend97y}
||g^\var_0||_{H^3(\R^3 \times \R^3)} \leqslant C N  ||f^\var_0||_{H^3 (\R^3 \times \R^3)},
\end{equation}
where $C$ depends on the $L^{\infty}$ norm of $\chi_{M_0}$ and its derivatives,  (\ref{enumitemappend2})
follows as soon as we show that   $N=N(\var)=O(1)$ (see below).

So it is only left to check (\ref{enumitemappend4}).

Observe that
\begin{equation}\label{nbv}
\begin{array}{c}
|| f^{\varepsilon}_0 -  g^{\varepsilon_0}  ||_{L^2} =
|| f^{\varepsilon}_0 -  N\widetilde f^{\varepsilon}_0 \,  \chi_{M_0}||_{L^2} \leqslant \\ { } \\

\leqslant ||(1 -N \chi_{M_0}) f^\var_0||_{L^2} +  ||N \chi_{M_0}( f^\var_0-\widetilde f^\var_0)||_{L^2} \leqslant \\ { } \\

\leqslant |1-N|\,\,||f^\var_0||_{L^2} + N ||(1-\chi_{M_0})f^\var_0||_{L^2} + N||f^\var_0-\widetilde f^\var_0||_{L^2} 
\end{array}
\end{equation}
Obviously, the estimate for the second term of the last line of (\ref{nbv}) comes from assumption (\ref{truc}), 
and for the third term from the fact that $||f^\var_0-\widetilde f^\var_0||_{L^2}=O(\var\Vert f_0^\var\Vert_{H^2})$  as shown in lemma A.1 of \cite{AP}. To see that the first term is $O(\var^\alpha)$, i.e. that $|1-N|= O(\var^\alpha)$, one has to observe that (denote for brevity $\delta^\var(k)=\frac{1}{(\pi\var)^{\frac{3}2}} e^{-\frac{|k|^2}{\var}}$)
\begin{equation} \label{eqdefNappend}
\begin{array}{c}
 |1-\frac{1}{N}| =| \int{(\widetilde f^\var_0\chi_{M_0}-\widetilde f^\var_0) dxdk}|\leqslant 
||(1-\chi_{M_0}) \widetilde{f}^\var_0   ||_{L^1}  \leqslant \\ { } \\

\leqslant \int\limits_{|k|>\frac{M_0}{2}} { \delta^\var(k-k')f^\var_0(x',k')dx'dk'dk }=
 \int\limits_{\begin{scriptsize}
\begin{array}{c} 
 |k|>\frac{M_0}{2} \\
 |k'|>\frac{M_0}{4}
\end{array} \end{scriptsize}
 } { \delta^\var(k-k')f^\var_0(x',k')dx'dk'dk } + \\ { } \\
 
+ \int\limits_{\begin{scriptsize}
\begin{array}{c} 
 |k|>\frac{M_0}{2} \\
 |k'|<\frac{M_0}{4}
\end{array} \end{scriptsize}
} { \delta^\var(k-k')f^\var_0(x',k')dx'dk'dk } \leqslant \int\limits_{k \in \mathbb{R}^3}{\delta^\var(k)dk } \, \int\limits_{|k'|>\frac{M_0}{4}} {|f^\var_0(x',k')|dx'dk'} + \\ { } \\

+\mathop{\sup} \limits_{\begin{scriptsize}
\begin{array}{c} 
 |k|>\frac{M_0}{2} \\
 |k'|<\frac{M_0}{4}
\end{array} \end{scriptsize}} |\delta^\var(k-k')| \,\,\,||f^\var_0||_{L^1} =
O(\var^\al)
\end{array}
\end{equation}
where we made use of assumptions (\ref{truc22}) and (\ref{truc23}) in the final step.

Now obviously,
\begin{equation}
\begin{array}{c}
\left|{1-\frac{1}{N} }\right| \leqslant C\var^\al \,\, \Rightarrow \,\, |N-1| \leqslant C\var^\al   |N| \,\, \Rightarrow \,\,\\ { } \\
\Rightarrow \,\,
 |N-1| \leqslant C \var^\al  (1 +|N-1|) \,\, \Rightarrow \,\, |N-1| \leqslant \frac{C \var^\al }{1-C \var^\al }=O(\var^\al ).
 \end{array}
\end{equation}

\end{appendix}

\vskip 0.5cm

\noindent \textbf{Acknowledgments}\\
The work of F. Pezzotti has been partially supported by Project CBDif-Fr ANR-08-BLAN-0333-01.

\end{document}